\documentstyle[aps,prb,multicol]{revtex}

\begin{document}

\twocolumn[
\hsize\textwidth\columnwidth\hsize\csname@twocolumnfalse\endcsname

\title{Effective-field-theory Approach to Persistent Currents}

\author{H. J. Bussemaker and T. R. Kirkpatrick}

\address{Institute for Physical Science and Technology 
	 and Department of Physics and Astronomy\\
         University of Maryland, College Park Maryland 20742\\
	 (Phys.\ Rev.\ B 56, 4529--4540 (1997).)}

\maketitle

\begin{abstract}
Using an effective-field-theory (nonlinear $\sigma$ model) description of 
interacting electrons in a disordered metal ring enclosing magnetic flux, 
we calculate the moments of the persistent current distribution, in terms 
of interacting Goldstone modes (diffusons and cooperons).
At the lowest or Gaussian order we reproduce well-known results for the average
current and its variance that were originally obtained using diagrammatic 
perturbation theory.
At this level of approximation the current distribution can be shown to be 
strictly Gaussian.
The nonlinear $\sigma$ model provides a systematic way of calculating higher-order 
contributions to the current moments.
An explicit calculation for the average current of the first term beyond 
Gaussian order shows that it is small compared to the Gaussian result;
an order-of-magnitude estimation indicates that the same is true for all 
higher-order contributions to the average current and its variance.
We therefore conclude that the experimentally observed magnitude of 
persistent currents cannot be explained in terms of interacting diffusons 
and cooperons.

\end{abstract}
]

\newpage

\section{Introduction}

The unexplained large magnitude of observed persistent currents in disordered
gold and copper rings --- with a size on the order of micrometers and at temperatures 
in the milli-Kelvin range --- has been an outstanding problem in condensed matter 
physics over the past half decade.
The phenomenon was already predicted by Hund.\cite{Hund}
Due to the high degree of difficulty of the experiments there is only a very 
limited number of observations that have been published.
\cite{Mailly95,Levy90,Chandrasekhar91,Mohanty96}
In the case of clean GaAs rings, where the electrons move ballistically,
the experiment \cite{Mailly95} is in good agreement with theory.\cite{clean-theory}
When impurities are present however, the electrons move diffusively;
in this case there is a discrepancy of one or two orders of magnitude between the
experiments \cite{Levy90,Chandrasekhar91} and the theoretical predictions,
especially in the case of the single-ring experiment.\cite{Chandrasekhar91}

On the theoretical side there has recently been a lot of activity.
\cite{exact-diag,hartree-fock,Bouchiat89,Cheung89,Ambegaokar91,Eckern91,%
Eckern92,Smith92,Beal-Monod92,Kopietz93,Vignale94,Mueller94,%
Schmid91,VonOppen91,Altshuler91a}
However, in spite of all efforts, there is still no satisfactory 
and widely accepted explanation for the observed current magnitude.
It seems that a resolution of the problem will necessarily involve taking into
account the (screened) Coulomb interaction between electrons.
This belief is based on the results of exact diagonalization studies for
tight-binding models, showing that electron-electron interaction can 
significantly enhance the magnitude of the current.\cite{exact-diag}
The size of the systems that can be studied in this manner is 
however far to small for any firm conclusion to be drawn from it.
Larger system sizes can be achieved by numerically solving the problem in
Hartree-Fock approximation, a method that has recently become popular.
\cite{hartree-fock}
This approach however neglects potentially important electronic correlations,
and should therefore be treated with some suspicion.

Analytic approaches are mostly based on Green's function techniques for
disordered systems;\cite{Abrikosov}
a nice summary of the work performed along these lines can be found in 
Ref.~\onlinecite{Smith-thesis}.
Two widely accepted results for the average\cite{Ambegaokar91} and
typical\cite{Cheung89} current
are based on these methods, and involve the summation of an infinite series of
ladder diagrams.  These resummations give rise to so-called diffusons,
related to the classical diffusion mode associated with the conserved electron 
density, and cooperons, the time reversed equivalent of diffusons.
Unfortunately, the results obtained along this route do not explain the experiments.
At higher orders in perturbation theory it becomes very difficult to determine 
which classes of diagrams are important, and there has been some controversy 
on this subject.\cite{Eckern92,Smith92,Beal-Monod92}
It is clear that a resolution of the question whether or not a theory in terms
of interacting diffusons and cooperons is capable of explaining the experimental
observations, requires an unambiguous method for generalizing the known results
to higher orders in perturbation theory.

A consistent theoretical description of the interplay between disorder 
and electron-electron interaction is a notoriously hard problem 
in solid state physics.
In the context of localization, a mapping of the quantum many-body problem 
to a classical field theory 
--- a matrix nonlinear $\sigma$ model that provides a systematic description 
in terms of the slow Goldstone modes associated with a continuous symmetry of
the order parameter --- has proved very useful.
The method was pioneered by Wegner,\cite{Wegner} and extended to include
electron-electron interactions by Finkel'stein.\cite{Finkelstein}
A general and self-contained derivation can be found in a recent review by 
Belitz and Kirkpatrick.\cite{revmodphys}

The nonlinear $\sigma$ model provides the starting point for the theory presented 
in this paper.  Here we will only give an outline of the derivation,
but we will provide full details when appropriate, in particular since there are
necessarily some technical differences with Ref.~\onlinecite{revmodphys}.
By analogy with the localization problem we expect that the nonlinear $\sigma$ model
will allow us to perform a perturbative analysis in a much more efficient 
manner than is possible using standard perturbation theory.
The results however will be equivalent, and in principle one should be able to
identify the corresponding classes of Feynman diagrams.
This has been done explicitly in some cases for the localization problem.

The outline of the paper is as follows.
In Sec.~\ref{sec:formulation} we formulate the problem.
Subsequently , in Sec.~\ref{sec:sigma-model}, we introduce the nonlinear 
$\sigma$ model and present a systematic method for calculating moments of the 
current distribution.
In Sec.~\ref{sec:Gaussian} we evaluate all moments at the lowest or Gaussian order,
and show that at this level of approximation the current distribution is strictly
Gaussian.
In Sec.~\ref{sec:beyond-Gaussian} the first contribution to the average current
beyond Gaussian order is calculated in full detail; an estimate is obtained for 
the magnitude of all higher-order contributions to the first and second moment of 
the current distribution;
furthermore it is argued that a low-frequency divergence occurs at all orders 
beyond Gaussian.
The paper ends with a discussion in Sec.~\ref{sec:discussion}.

\section{Formulation of the problem}
\label{sec:formulation}

Let us describe in detail the problem we consider.
We will use units in which $\hbar=c=1$.
The circumference of the ring is denoted by $L$ and its transverse width
by $L_\bot$.
When the ring is placed in a stationary magnetic field, an equilibrium 
current,\cite{clean-theory}
\begin{equation}
	I = -\frac{\partial\Omega}{\partial\phi},
\end{equation}
will flow in response along the circumference of the ring.
Here $\Omega$ is the free energy and
\begin{equation}
	\phi = \oint {\bf A}\cdot{\bf dl}
\end{equation}
is the magnetic flux enclosed by the ring.
The density of impurities is specified in terms of the mean free path $\ell$,
which is large compared to the Fermi wavelength: $k_F \ell \sim 100$.
The width $L_\perp\sim\ell$ of the ring is much smaller than its circumference $L$;
typically we have $L/\ell\sim 100$.
Thus it is reasonable to assume that we can describe the behavior of the electrons
in terms of slow diffusive modes.

It is well known that in strictly one-dimensional systems the electrons
are localized on a scale $\xi\sim\ell$.
In a quasi-one-dimensional geometry however, the localization occurs on a scale 
$\xi\sim(k_{\!F}L_\bot)^2\ell \gg \ell$, where $(k_{\!F}L_{\bot})^2$
is the number of transverse channels.
Since $\xi\gg L$ we expect the electrons to behave diffusively.\cite{Thouless}

In a clean ring we expect the persistent current to be on the order of
the {\em ballistic} current scale,
\begin{equation}\label{I_B}
	I_B = \frac{e}{t_B} = \frac{e v_F}{L},
\end{equation}
where $t_B=L/v_F$ is the time it takes an electron to travel around the ring at 
the Fermi velocity $v_F$.
When disorder is present the electron motion is diffusive, with an associated
time scale $t_D = L^2/D$, where $D=v_F\ell/d$ is the diffusion coefficient
and $d=3$ the dimensionality.
Thus the {\em diffusive} current scale is
\begin{equation}\label{I_D}
	I_D = \frac{e}{t_D} = \Big(\frac{e v_F}{L}\Big)
	\Big(\frac{\ell}{L}\Big) \ll I_B.
\end{equation}

Three experiments have been performed on disordered metal rings:
(i) a multiring experiment\cite{Levy90} in which the average current per 
ring in an array of $10^7$ rings was found to be on the order of $I_D$;
(ii) a single-ring experiment \cite{Chandrasekhar91} in which much larger 
currents on the order of $I_B$ were observed;
(iii) a 30-ring experiment \cite{Mohanty96} that interpolates between the
first two experiments.
The second experiment in particular poses a challenge, since theory so far predicts
a value for the typical current that is one or two orders of magnitude smaller
than what has been observed experimentally.

\section{Nonlinear $\sigma$ model approach}
\label{sec:sigma-model}

\subsection{The model}

A field-theoretic description of interacting electrons in a random
potential starts from an expression for the partition function $Z$ in
terms of a functional integral of the form
\begin{equation}
	Z = \int D[\bar\psi,\psi] e^{S[\bar\psi,\psi]}.
\end{equation}
Here $\psi_\sigma({\bf x},\tau)$ is a field that depends on
the spin label $\sigma=\pm\case{1}{2}$, position ${\bf x}$, and imaginary 
time $\tau$, 
while $\bar\psi_\sigma({\bf x},\tau)$ denotes the conjugate field.
The action $S[\bar\psi,\psi]$ includes both the disorder potential and the 
repulsive electron-electron interaction.
Since we are dealing with a fermionic system, $\psi$ and $\bar\psi$ are 
anticommuting (Grassmann) fields; their Fourier transform is defined by
\begin{mathletters}
\begin{equation}
	\psi_\sigma({\bf x},\tau) = 
	\sqrt{\frac{T}{LL_\bot^2}}\
	\sum_{{\bf k},n} e^{i({\bf k}\cdot{\bf x}-\omega_n\tau)}
	\psi_\sigma({\bf k},\omega_n);
\end{equation}
\begin{equation}
	\bar\psi_\sigma({\bf x},\tau) = 
	\sqrt{\frac{T}{LL_\bot^2}}\
	\sum_{{\bf k},n} e^{-i({\bf k}\cdot{\bf x}-\omega_n\tau)} 
	\bar\psi_\sigma({\bf k},\omega_n),
\end{equation}
\end{mathletters}
where $\omega_n=2\pi(n+\case{1}{2})$, with $n$ integer, is a fermionic Matsubara
frequency.
The free energy is given by
\begin{equation}
	\Omega = -T \langle\ln Z\rangle,
\end{equation}
where $\langle\cdots\rangle$ denotes an average over disorder
and $T$ is the temperature 
(we use units in which Boltzmann's constant $k_{\rm B}\equiv1$).
To evaluate $\langle\ln Z\rangle$ we use the replica trick,
which makes use of the identity
\begin{equation}\label{replica-trick}
	\ln Z = \lim_{N\to0} \frac{1}{N}(Z^N-1),
\end{equation}
where the partition function
\begin{equation}
	Z^N = \int \prod_{\alpha=1}^N D[\bar\psi^\alpha,\psi^\alpha] 
	\exp\left[\sum_{\alpha=1}^N S[\bar\psi^\alpha,\psi^\alpha]\right].
\end{equation}
describes $N$ identical ``replicas'' of the original system.
Averaging over a Gaussian disorder distribution introduces quartic interaction
terms between different replicas;
these terms occur in addition to single-replica quartic terms describing 
electron-electron interaction.
Subsequently, all these four-fermion terms are decoupled by performing 
Hubbard-Stratonovich transformations. 
This introduces a classical matrix field $Q^{\alpha\beta}_{nm}$
with replica labels $\alpha,\beta$ and Matsubara frequency labels $n,m$.
The resulting action is quadratic in the fermionic degrees of freedom
$\psi$ and $\bar\psi$, so that these can be integrated out.
The result is a classical field theory of the form:
\begin{equation}\label{Q-pathintegral}
	Z^N = \int D[Q] e^{S[Q]}.
\end{equation}

The action $S[Q]$ has a continuous symmetry that was first recognized 
for noninteracting electrons by Wegner.\cite{Wegner}
In the low frequency limit, the Goldstone modes associated with this 
symmetry correspond to particle-hole excitations (``diffusons'') with 
$Q^{\alpha\beta}_{nm}$ corresponding to the composite variable
$\bar\psi^\alpha_n\psi^\beta_m$ or $\psi^\alpha_n\bar\psi^\beta_m$, 
and particle-particle excitations (``cooperons'') with 
$Q^{\alpha\beta}_{nm}$ corresponding to
$\psi^\alpha_n\psi^\beta_m$ or $\bar\psi^\alpha_n\bar\psi^\beta_m$.

The field $Q^{\alpha\beta}_{nm}({\bf x})$ is an element in four-dimensional
bispinor space, spanned by the charge (particle/hole) and spin (up/down)
degrees of freedom.
A suitable basis in this space is formed by the matrices
$\epsilon_r \otimes s_i$, where $r,i=0,\ldots,3$.
The matrices
\[
	~
	\epsilon_0 \!=\! \left({\!\!
	\begin{array}{cc} 1&0 \\ 0&0 \end{array}}\!\!\right)
	\!,\hfill
	\epsilon_1 \!=\! \left({\!\!
	\begin{array}{cc} 0&1 \\ 0&0 \end{array}}\!\!\right)
	\!,\hfill
	\epsilon_2 \!=\! \left({\!\!
	\begin{array}{cc} 0&0 \\ 1&0 \end{array}}\!\!\right)
	\!,\hfill
	\epsilon_3 \!=\! \left({\!\!
	\begin{array}{cc} 0&0 \\ 0&1 \end{array}}\!\!\right)
	\!.
	~
\]
span the charge space. 
Note that this definition is different from the one used in 
Ref.~\onlinecite{revmodphys}, since the latter is not useful for taking
into account the effect of the magnetic flux enclosed by the ring 
(see Sec.~\ref{sec:fluxes-wavenumbers} below).
This modification gives rise to a somewhat different formalism at a technical 
level; we will explicitly indicate differences as they occur.
For the spin space we use the same representation as in Ref.~\onlinecite{revmodphys}.
It is spanned by
\[
	~
	s_0 \!=\! \left({\!\!
	\begin{array}{cc} 1&0 \\ 0& 1 \end{array}}\!\!\right)
	\!,\hfill
	s_1 \!=\! \left({\!\!
	\begin{array}{cc} 0&i \\ i& 0 \end{array}}\!\!\right)
	\!,\hfill
	s_2 \!=\! \left({\!\!\!
	\begin{array}{rr} 0&1 \\-1& 0 \end{array}}\!\!\right)
	\!,\hfill
	s_3 \!=\! \left({\!\!
	\begin{array}{rr} i&0 \\ 0&-i \end{array}}\!\!\right)
	\!.
	~
\]
In terms of this basis, $Q^{\alpha\beta}_{nm}$ is given by
\begin{equation}\label{bispinor-basis}
	Q^{\alpha\beta}_{nm} = \sum_{r=0}^3 \sum_{i=0}^3
	(\epsilon_r \otimes s_i)\ {}^i_r\!Q^{\alpha\beta}_{nm}.
\end{equation}
The matrix $Q$ can be chosen to be Hermitian: 
\begin{equation}\label{hermitian}
	Q=Q^\dagger.
\end{equation}
In addition, it is required that as an operator in bispinor space,
$Q$ satisfy the condition
\begin{equation}\label{charge-conj}
	C^T Q^T C = Q,
\end{equation}
where $C$ is the charge conjugation operator, given by
\begin{equation}
	C = \left(\begin{array}{cc} 0 & s_2 \\ s_2 & 0 \end{array}\right)
	= i (\epsilon_0+\epsilon_3) \otimes s_2.
\end{equation}
It follows from Eqs.~(\ref{hermitian}) and (\ref{charge-conj}) that
\begin{equation}\label{symprop}
	({}^i_0 {Q^{\alpha\beta}_{nm}})^\star = {}^i_3 Q^{\alpha\beta}_{nm},
	\qquad\quad
	({}^i_1 {Q^{\alpha\beta}_{nm}})^\star = {}^i_2 Q^{\alpha\beta}_{nm}.
	\nonumber
\end{equation}
We expand around the ``Fermi-liquid'' saddle point\cite{revmodphys}
\begin{equation}
	(Q_{\rm SP})^{\alpha\beta}_{nm} = [(\epsilon_0+\epsilon_3)\otimes s_2]
	\ \delta_{\alpha\beta} \delta_{nm}\ {\rm sgn}(\omega_n).
\end{equation}
An alternative notation for $Q_{\rm SP}$ is
\begin{equation}
	Q_{\rm SP} = \left(\begin{array}{cc}1 & 0\\ 0 & -1 \end{array}\right),
\end{equation}
where the $2\times2$ matrix structure refers to the Matsubara frequency
indices in the following way:
\begin{equation}
	\left(\begin{array}{ccc}
	n,m\geq0 & \quad & n\geq0,\ m<0 \\ &&\\
	n<0,\ m\geq0 & \quad & n,m<0
	\end{array}\right).
\end{equation}
In addition to the slow Goldstone modes there are massive modes that can 
effectively be integrated out by imposing the constraints
\begin{equation}
	Q^2 = 1; \qquad\quad {\rm tr}\ Q=0.
\end{equation}
These constraints can be eliminated by choosing a suitable parametrization:
we will consider fluctuations
\begin{equation}
	\widetilde Q=Q-Q_{\rm SP},
\end{equation} 
of the form
\begin{equation}\label{parametrization}
	\widetilde Q = \left(\begin{array}{ccc}
		\sqrt{1-qq^\dagger}-1 & \quad & q \\ &&\\
		q^\dagger & \quad & -\sqrt{1-q^\dagger q}+1 \\
	\end{array}\right),
\end{equation}
where the unconstrained field $q^{\alpha\beta}_{nm}({\bf x})$ has 
$n\geq0$, $m<0$; it can again be expanded in terms of
$(\epsilon_r\otimes s_i)$ in exactly the same way as $Q$
[see Eq.~(\ref{bispinor-basis})].
Expanding the action $S[\widetilde Q]$ in terms of the gradient operator and 
frequency, and neglecting terms of order $\nabla^4$, $\omega^2$, and 
$\nabla^2 \omega$, we obtain the action
\begin{eqnarray}
	S[Q] &=& \frac{-1}{2G} \int d{\bf x}\ {\rm tr}(\nabla 
			\widetilde Q({\bf x}))^2
	\nonumber\\ && \mbox{} 
	       + 2H \int d{\bf x}\ {\rm tr}(\Omega \widetilde Q({\bf x}))
	       + S_{\rm int}[Q],
\end{eqnarray}
where
\begin{equation}
	\Omega^{\alpha\beta}_{nm} = \delta_{\alpha\beta} \delta_{nm}
	\omega_n (\epsilon_0+\epsilon_3) \otimes s_0.
\end{equation}
The coupling constants $G$ and $H$ are defined as
\begin{equation}
	G=\frac{4}{\pi N_{\!F} D}, 
	\qquad
	H=\frac{\pi N_{\!F}}{4},
\end{equation}
where $N_{\!F}$ denotes the saddle point density of states at the Fermi energy.
Note that $GH=1/D$.
The interaction term $S_{\rm int}[Q]$ term is defined in 
Ref.~\onlinecite{revmodphys}.
It is quadratic in $\widetilde Q$ and consists of three parts, with coupling constants
$K_{\rm s}$ for the singlet channel, $K_{\rm t}$ for the triplet channel,
and $K_{\rm c}$ for the singlet Cooper channel;
the coupling constant for the triplet Cooper channel vanishes as a consequence
of the Fermi exclusion principle.

Using the parametrization (\ref{parametrization}) we can write 
the action in terms of an expansion in powers of $q$ as
\begin{equation}
	S[q] = \sum_{n=2}^\infty S^{(\!n\!)}[q].
\end{equation}
It should be stressed that although this is an expansion in the strength of
the disorder, $1/k_{\!F}\ell$, the interaction coupling constants
$K_{\rm s,t,c}$ are taken into account to all orders for each term in the 
expansion.

Higher order gradient terms can be added to the action in a systematic
manner.\cite{Altshuler91b}
These terms can be shown give contributions to the persistent current
that are smaller than the $O(\nabla^2)$ result by a factor $(\ell/L)^2$, 
and they can consequently be neglected.

When the behavior of the coupling constants under renormalization group (RG)
transformations is considered, the interaction constant may under certain
circumstances be strongly renormalized to large 
values.\cite{Finkelstein,revmodphys}
However, since here we are dealing with mesoscopic systems, in which the
finite system size provides an upper bound on the spatial rescaling underlying 
the RG flow equations, these renormalization effects will be unimportant for 
our purposes.

\subsection{Fluxes and replicas}

The key ingredient in the calculation the $n$th moment of the current 
distribution, $\langle I^n(\phi)\rangle$, is the introduction 
of $n$ classes of replicas, each having their own independent flux $\phi_i$.
Let the total number of replicas be $N=\sum_{i=1}^n N_i$,
and let the fluxes be assigned as follows:
\begin{eqnarray}
	\alpha &=& \underbrace{1,2,\ldots,N_1}_{\phi_1},\ 
	\underbrace{(N_1+1),\ldots,(N_1+N_2)}_{\phi_2},\ \ldots\ , 
	\nonumber\\
	&& \mbox{} 
	\underbrace{(N_1+\dots+N_{n-1}+1),\ldots,(N_1+\ldots+N_n)}_{\phi_n}.
\end{eqnarray}
Using the identity
\begin{equation}
	\prod_{i=1}^n Z^{N_i}(\phi_i)
	= \prod_{i=1}^n \left[1+N_i\ln Z(\phi_i)+O(N_i^2)\right],
\end{equation}
we derive that 
\begin{eqnarray}
	\frac{\langle I^n(\phi)\rangle}{(-T)^n} &=&
	\lim_{N_1\to0} \cdots \lim_{N_n\to0}
	\left(\prod_{i=1}^n\frac{1}{N_i}\right)
	\left(\prod_{i=1}^n\frac{\partial}{\partial\phi_i}\right)
	\nonumber\\ && \mbox{} \times
	\left.\left\langle Z^{N_1+\cdots+N_n}(\phi_1,\ldots,\phi_n)
	\right\rangle\right|_{\phi_i=\phi}.
\end{eqnarray}
Using the classical path integral representation for
$\langle Z^N\rangle$ given in Eq.~(\ref{Q-pathintegral})
we thus have for the first and second moment:
\begin{mathletters}
\begin{equation}
	\frac{\langle I(\phi)\rangle}{T} =
	- \int{\cal D}[Q]\ e^{S[Q]}\ 
	\left(\frac{\partial S[Q]}{\partial\phi}\right);
\end{equation}
\begin{eqnarray}\label{Isq}
	\frac{\langle I^2(\phi)\rangle}{T^2} &=&
	\int{\cal D}[Q]\ e^{S[Q]}
	\nonumber\\
	&& \hspace{-30pt} \times \left[
	\left(\frac{\partial S[Q]}{\partial\phi}\right)^{\!2}
	+ \left( \frac{\partial^2 S[Q]}
	{\partial\phi_1\,\partial\phi_2}\right)_{\phi_{1,2}=\phi}
	\right].
\end{eqnarray}
\end{mathletters}

\subsection{Fluxes and wave numbers}
\label{sec:fluxes-wavenumbers}

Next we turn to the question of which wave numbers are allowed.  
Since the width of the ring is much smaller than its circumference,
$L_\bot\ll L$, we neglect excitations in the transverse direction,
i.e.\ we only consider ${\bf k}=(k,0,0)$.
At low temperatures in particular, this ``freezing out'' of the transverse 
degrees of freedom is justified since transverse excitations have a much
higher energy than longitudinal ones.

We furthermore assume that the vector potential is constant over the 
circumference of the ring.  
We can then `gauge away' the magnetic field, and summarize the effect 
of the magnetic field by means of the twisted boundary condition
\begin{equation}
	\psi(x+L) = e^{2\pi i\phi/\phi_0} \psi(x),
\end{equation}
where $\phi_0=2\pi/e$ is the elementary flux quantum, in units where $\hbar=c=1$.
Consequently the allowed wave numbers, with $n_k$ an integer, are
\begin{mathletters}
\begin{equation}
	\psi(k): \quad
	k = \frac{2\pi}{L}\left(n_k+\frac{\phi}{\phi_0}\right);
\end{equation}
\begin{equation}
	\bar\psi(k): \quad
	k = \frac{2\pi}{L}\left(n_k-\frac{\phi}{\phi_0}\right).
\end{equation}
\end{mathletters}
It follows from the definition of $\epsilon_r$ 
that for ${}^i_r\!Q^{\alpha\beta}_{nm}({\bf p})$ with ${\bf p}=(p,0,0)$
the allowed wave numbers $p$, with $n_p$ an integer, are
\begin{equation}\label{p}
	p = \frac{2\pi}{L}
	\left(n_p +
	\left[\begin{array}{c}-\\-\\+\\+\end{array}\right]_r 
	\left(\frac{\phi_\alpha}{\phi_0}\right)
	+
	\left[\begin{array}{c}+\\-\\+\\-\end{array}\right]_r 
	\left(\frac{\phi_\beta}{\phi_0}\right)\right).
\end{equation}
Note that for diffusons ($r=0,3$) the wave number $p$ 
depends on $(\phi_\alpha-\phi_\beta)$ while in the case of cooperons 
($r=1,2$) we have a dependence on $(\phi_\alpha+\phi_\beta)$.

\subsection{The zero-momentum mode}
\label{sec:zero-mode}

When a field theory in a finite geometry is considered, the zero-momentum 
mode usually requires special treatment.\cite{Zinn-Justin}
In the case of the nonlinear $\sigma$ model the integration over the zero-mode
corresponds to an average over the group of transformations that leave the
action invariant.
The moments of the current distribution are invariant under this group.
However, they are calculated by means of an expansion around a particular 
saddle point that breaks the symmetry of the group.
It is for this reason that the zero mode should be omitted in all expressions.

As will become clear later on, the expressions for the current moments will
in general contain sums over discrete momenta that are shifted by the
flux $\phi$ in the case of cooperons, while there is no dependence on $\phi$
for diffusons.
For example, when in Eq.~(\ref{p}) we choose $\phi_\alpha=\phi_\beta=\phi$ the
values of $p$ depend on $\phi$ for $r=1,2$ while they are independent of $\phi$
for $r=0,3$.
Therefore only in the case of diffusons does the zero-mode $p=0$ occur, 
and should it be omitted.
Later on, in Sec.~\ref{sec:sums} we will come back to the question what 
happens in the case of cooperons
when the flux $\phi$ is close to a multiple of $\phi_0$.

\section{Gaussian theory}
\label{sec:Gaussian}
\subsection{The Gaussian propagator}

To quadratic order in $q$ the action is given by the Gaussian part only,
\begin{equation}
	S^{(\!2\!)}[q] = \left(\frac{-2}{GLL_\bot^2}\right)
	\sum_{\bf p} \sum_{r,i}\ {}^i_rq_{12}({\bf p})\
	{}^i_rM_{12,34}({\bf p})\ {}^i_rq^*_{34}({\bf p}),
\end{equation}
where $1\equiv\{\alpha_1,n_1\}$, etc.
In the case of diffusons ($r=0,3$), the matrix $M$ is given by
\begin{eqnarray}
	{}^i_{0,3}M_{12,34}({\bf p})
	&=& \delta_{\alpha_1\alpha_3}\ \delta_{\alpha_2\alpha_4}\
	\delta_{n_1-n_2,n_3-n_4}
	\nonumber\\ && \mbox{}\times
	\Big\{\delta_{n_1n_3}[p^2 + GH(\omega_{n_1}-\omega_{n_2})]
	\nonumber\\ && \mbox{} \hspace{20pt}
	+ \delta_{\alpha_1\alpha_2}\ 2\pi T G K_{s,t}\Big\},
\end{eqnarray}
where the interaction term contains the singlet coupling constant $K_s$ 
for $i=0$ and the triplet coupling constant $K_t$ for $i=1,2,3$.
For cooperons ($r=1,2$) we have
\begin{eqnarray}
	{}^i_{1,2}M_{12,34}({\bf p})
	&=& \delta_{\alpha_1\alpha_3}\ \delta_{\alpha_2\alpha_4}\
	    \delta_{n_1+n_2,n_3+n_4}
	\nonumber\\ && \mbox{}\times
	\Big\{\delta_{n_1n_3}[p^2 + GH(\omega_{n_1}-\omega_{n_2})]
	\nonumber\\ && \mbox{} \hspace{20pt}
	+\delta_{i,0}\ \delta_{\alpha_1\alpha_2}\ 2\pi T G K_{\rm c} \Big\},
\end{eqnarray}
where $K_c$ is the cooperon coupling constant.

The Gaussian propagator is given by
\begin{equation}
	\langle{}^i_rq_{12}({\bf p}_1) {}^j_sq_{34}^\star({\bf p}_2)\rangle
	= \frac{GLL_\bot^2}{4}\ \delta_{rs}\ \delta_{ij}\ 
	\delta_{{\bf p}_1,{\bf p}_2} {}^i_rM^{-1}_{12,34}({\bf p}_1).
\end{equation}
Note the slight difference with Ref.~\onlinecite{revmodphys}.
Another modification is that due to the symmetry property (\ref{symprop})
we have
\begin{eqnarray}
	\langle{}^i_rq_{12}({\bf p}_1) {}^j_{3-s}q_{34}({\bf p}_2)\rangle
	&=& \langle{}^i_rq_{12}({\bf p}_1) {}^j_sq_{34}^\star({\bf p}_2)\rangle
	\nonumber\\
	&=& \langle{}^i_{3-r}q_{12}^\star({\bf p}_1) 
	           {}^j_sq_{34}^\star({\bf p}_2)\rangle
\end{eqnarray}
The inverse $M^{-1}$ of the matrix $M$ has been calculated in 
Ref.~\onlinecite{revmodphys}.
We cite the result, in terms of the diffusive propagators
\begin{eqnarray}
	{\cal D}_n(p) &=& \left[ p^2 + GH\Omega_n\right]^{-1}, \\
	{\cal D}_n^{\rm s,t}(p) &=& 
	\left[ p^2 + G(H+K_{\rm s,t})\Omega_n\right]^{-1}, 
	\nonumber\\
	\Delta{\cal D}_n^{\rm s,t}(p) &=&
	{\cal D}_n^{\rm s,t}(p) - {\cal D}_n(p).
\end{eqnarray}
For diffusons we have
\begin{eqnarray}
	{}^i_{0,3}M^{-1}_{12,34}({\bf p})
	&=& \delta_{\alpha_1\alpha_3}\ \delta_{\alpha_2\alpha_4}\
	\delta_{n_1-n_2,n_3-n_4}
	\nonumber\\ && \hspace{-70pt}\mbox{}\times
	\left\{\delta_{n_1n_3}\ {\cal D}_{n_1-n_2}({\bf p})
	+ \frac{\delta_{\alpha_1\alpha_2}}{n_1-n_2}\ 
	\Delta{\cal D}^{\rm s,t}_{n_3-n_4}({\bf p}) \right\},
\end{eqnarray}
while for cooperons
\begin{eqnarray}
	{}^i_{1,2}M^{-1}_{12,34}({\bf p})
	&=& \delta_{\alpha_1\alpha_3}\ \delta_{\alpha_2\alpha_4}\
	\delta_{n_1+n_2,n_3+n_4}
	\nonumber\\ && \hspace{-40pt}\mbox{}\times
	\Biggl\{ \delta_{n_1n_3}\ {\cal D}_{n_1-n_2}({\bf p})
	- {\cal D}_{n_1-n_2}({\bf p})\ {\cal D}_{n_3-n_4}({\bf p})
	\nonumber\\ && \hspace{-10pt} \mbox{}\times
	\frac{\delta_{i,0}\ \delta_{\alpha_1\alpha_2}\ 2\pi T G K_{\rm c}}
	{1+2\pi T G K_{\rm c} f_{n_1+n_2}({\bf p})} \Biggr\},
\end{eqnarray}
where
\begin{equation}
	f_{n_1+n_2}({\bf p}) = \sum_{n_1,n_2} \delta_{n,n_1+n_2}
	{\cal D}_{n_1+n_2}({\bf p}).
\end{equation}
Note that contributions related to electron-electron interactions 
(i.e., proportional to $K_{\rm s,t,c}$) require that all replicas be the same:
$\alpha_1=\alpha_2=\alpha_3=\alpha_4$.

\subsection{Logarithmic suppression of $K_{\rm c}$}
\label{sec:log-suppression}

The interaction part of the cooperon propagator (i.e., the part proportional to 
$K_{\rm c}$) is suppressed by a factor 
$[1+2\pi T G K_{\rm c} f_{n_1+n_2}({\bf p})]^{-1}$.
It is important to estimate the quantitative effect of this term.
The term $f_n(p)$ diverges logarithmically, as
\begin{eqnarray}
	2\pi TGK_{\rm c} f_n(p)
	&=& 8\pi TN_{\!F}\bar V
	\sum_{m=0}^\infty \frac{1}{D p^2 + 2\pi T(n+2m)} 
	\nonumber\\ 
	&\simeq& 2N_{\!F}\bar V
	\int_T^{1/\tau} d\omega_m \frac{1}{Dp^2 + \omega_n + 2\omega_m} 
	\nonumber\\
	&=& 2N_{\!F}\bar V \ln\left[
	\frac{Dp^2+\omega_n + \tau^{-1}}{Dp^2+\omega_n + T} \right].
\end{eqnarray}
The upper and lower bounds on the frequency integral are provided by the
average collision frequency and the temperature, respectively.

Since the electron-electron interactions are screened and therefore short ranged,
they can be approximated by a point interaction with strength $\bar V$.
From the definition of $K_{\rm c}$ in Ref.~\onlinecite{revmodphys} it follows
that $K_{\rm c}=\pi N_{\!F}^2\bar V/2$, or $GK_{\rm c}=2N_{\!F}\bar V/D$.
A typical value\cite{Ambegaokar91} for the electron-electron interaction is 
$N_{\!F}\bar V\simeq0.3$. 

The dominant contribution to the persistent current comes from wave numbers and
frequencies on diffusive scales, $p\sim2\pi/L$ and $\omega_n\sim D(2\pi/L)^2$,
and $T\lesssim T_0$ with $T_0=D/L^2$, so that the argument of the logarithm has magnitude
$(L/2\pi)^2/D\tau = 3(L/\ell)^2/(2\pi)^2 \gg 1$.
Using $N_{\!F}\bar V\simeq 0.3$ and $L/\ell\simeq 140$ we estimate 
that $[1+2\pi TGK_{\rm c} f_n(p)] \simeq 5$, in agreement with Eckern,\cite{Eckern91}
who obtained the same result by summing an infinite series of Feynman diagrams.

In the remainder of the paper we will for each occurrence of the cooperon propagator
replace $N_{\!F}\bar V$ by its effective 
value $N_{\!F}\bar V_{\rm eff}\simeq 0.2N_{\!F}\bar V \simeq 0.06$, and drop the 
denominator $[1+2\pi TGK_{\rm c}f({\bf p})]$.

It is worth noting the close connection between the logarithmic suppression
described here and the logarithm appearing in BCS theory.  Both result
from the summation of an infinite series of diagrams involving multiple 
interactions.
In the case of a phonon-mediated attractive electron-electron interaction, 
there is an enhancement rather than a suppression of $K_{\rm c}$, which for 
low enough temperatures leads to destabilization of the Fermi liquid ground 
state and the emergence of superconductivity.

\subsection{Calculation of $\langle I \rangle^{(2)}$}
\label{sec:I-Gaussian}

To calculate $\langle I\rangle^{(\!2\!)}$ in Gaussian approximation we
note that
\begin{equation}
	\frac{\partial S^{(\!2\!)}[q]}{\partial\phi} =
	\frac{-2}{GLL_\bot^2} \sum_{p} \sum_{12} \sum_{ri}
	(2p)\ \frac{\partial p}{\partial\phi}\
	{}^i_rq_{12}(p)\ {}^i_rq^*_{12}(p),
\end{equation}
where the allowed wave numbers are
\begin{equation}\label{p-phi}
	p = \frac{2\pi}{L} 
	\left(n_p + \left[\begin{array}{c}0\\-\\+\\0\end{array}\right]_r 
	\left(\frac{2\phi}{\phi_0}\right)\ \right).
\end{equation}
All $N=N_1$ replicas have flux $\phi$.
Due to presence of the factor $(\partial p/\partial\phi)$ only cooperons
($r=1,2$) contribute to the average current.
The result is
\begin{equation}\label{I}
	\langle I\rangle^{(\!2\!)} =
	\frac{T}{2}\sum_p \sum_{12} \sum_{r=1,2}
	(2p) \left(\frac{\partial p}{\partial\phi}\right) M^{-1}_{12,12}(p).
\end{equation}
In the replica limit ($N\to0$) only terms proportional to $N$ survive.
It follows that only the interacting part of $M^{-1}$ contributes, since it
is diagonal in the replica labels;
the noninteracting part of $M^{-1}$ yields a term proportional to $N^2$
and consequently vanishes in the replica limit.
The previous expression for the average current can be written as
\begin{equation}
	\langle I \rangle^{(\!2\!)} = 
	-\frac{\partial\langle\Omega(\phi)\rangle^{(\!2\!)}}{\partial\phi},
\end{equation}
with an average free energy given by
\begin{equation}
	\langle\Omega(\phi)\rangle^{(\!2\!)} =
	2 N_{\!F}\bar V T \sum_p \sum_{\omega>0} \frac{\omega}{Dp^2+\omega},
\end{equation}
where $\omega=2\pi Tn$, with integer $n\geq1$, 
and $p=(2\pi/L)$ $\times(n_p-2\phi/\phi_0)$.
This is exactly the Hartree-Fock result obtained in 
Ref.~\onlinecite{Ambegaokar91}, 
and the corresponding average current at $T=0$ is given by
\begin{equation}
	\langle I(\phi)\rangle^{(\!2\!)} \simeq
	I_D \left(\frac{8 N_{\!F}\bar V_{\rm eff}}{3\pi}\right)
	\sum_{m=1}^\infty \frac{\sin(4\pi m\phi/\phi_0)}{m^3},
\end{equation}
where $I_D$ is the diffusive current scale defined in Eq.~(\ref{I_D}) in
Sec.~\ref{sec:formulation}.
This result was obtained by exploiting the fact that $\langle I(\phi)\rangle$
is a periodic function of $\phi$ with period $\phi_0/2$.
In the calculation of the Fourier components the sum over wave numbers gets
replaced by an integral (see Ref.~\onlinecite{Ambegaokar91}).
An alternative calculation that does not rely on Fourier transformation of the
$\phi$ dependence shows that
\begin{equation}
	\sum_{m=1}^\infty \frac{\sin(4\pi m\phi/\phi_0)}{m^3} =
	\frac{1}{4} \int_0^\infty ds s^2 \frac{(\sinh s)(\sin 4\pi\phi)}
	{(\cosh s - \cos 4\pi\phi)^2}.
\end{equation}
Note that the integral exists for all $\phi$.

When the ``bare'' interaction $\bar V$ rather than the effective interaction
$\bar V_{\rm eff}$ is used, the ex\-pe\-ri\-mental 
\cite{Levy90} and theoretical \cite{Ambegaokar91} values for the 
average current $\langle I\rangle$ are in reasonable agreement.
The logarithmic suppression of the interaction discussed in 
Sec.~\ref{sec:log-suppression} severely worsens the agreement between 
theory and experiment, leading to a discrepancy by a factor of 5.

It is however perfectly consistent to keep the suppression term.
As was already noted earlier, the nonlinear $\sigma$ model provides an 
expansion in terms of the disorder strength, $1/k_{\!F}\ell$.
The inversion of the asymmetric matrix $M$ performed in the calculation
of the Gaussian propagator, however, amounts to an infinite resummation 
in terms of the interaction constant $K_{\rm c}$.
Therefore in a certain sense the electron-electron interaction is taken into
account exactly.

\subsection{Calculation of $\langle I^2 \rangle^{(2)}$}
\label{sec:Isq-Gaussian}

The second moment of the current distribution can be written as a sum
of a disconnected and a connected part,
\begin{equation}
	\langle I^2\rangle = \langle I\rangle^2 
	+ \langle I^2\rangle_c.
\end{equation}
According to Eq.~(\ref{Isq}) there are two contributions to the second
moment $\langle I^2\rangle^{(\!2\!)}$.
We first consider the term 
\begin{equation}
	\left\langle
	\frac{\partial^2 S^{(\!2\!)}[q]}{\partial\phi_1\partial\phi_2}
	\right\rangle^{(\!2\!)} =
	\frac{-4}{GLL_\bot^2} \sum_p \sum_{12} \sum_{r,i}
	\frac{\partial p}{\partial\phi_1} \frac{\partial p}{\partial\phi_2}
	{}^i_r M_{12,12}(p).
\end{equation}
Here the wave numbers are given by
\begin{equation}
	p = \frac{2\pi}{L}
	\left(n_p +
	\left[\begin{array}{c}-\\-\\+\\+\end{array}\right]_r 
	\left(\frac{\phi_{\alpha_1}}{\phi_0}\right)
	+
	\left[\begin{array}{c}+\\-\\+\\-\end{array}\right]_r 
	\left(\frac{\phi_{\alpha_2}}{\phi_0}\right)\right),
\end{equation}
so that $(\partial p/\partial\phi_1)(\partial p/\partial\phi_2)$
can only be nonvanishing if either
(i) $\phi_{\alpha_1}=\phi_1$ and $\phi_{\alpha_2}=\phi_2$, or 
(ii) $\phi_{\alpha_2}=\phi_1$ and $\phi_{\alpha_1}=\phi_2$.
In both cases we necessarily have $\alpha_1\neq\alpha_2$, so that
only the noninteracting part of $M^{-1}$ can contribute.
Furthermore it can easily be verified that in both cases the summation
over replica indices yields precisely the factor $N_1N_2$ that is needed 
for the term to survive in the replica limit.
Note that 
$(\partial p/\partial\phi_1)(\partial p/\partial\phi_2)=\pm e^2 (2\pi/L)^2$,
with a plus sign for diffusons ($r=0,3$) 
and a minus sign for cooperons ($r=1,2$).

A second contribution to $\langle I^2\rangle^{(\!2\!)}$ is of the type
\begin{eqnarray}
	\left\langle
	\frac{\partial S^{(\!2\!)}[q]}{\partial\phi_1}
	\frac{\partial S^{(\!2\!)}[q]}{\partial\phi_2}
	\right\rangle^{(\!2\!)} &\!\!=&
	\left(\frac{4}{GLL_\bot^2} \right)^2
	\sum_{p_1,p_2} \sum_{1234} \sum_{rs,ij}
	\frac{\partial p_1}{\partial\phi_1}
	\frac{\partial p_1}{\partial\phi_2}
	\nonumber\\ && \hspace{-90pt}\mbox{}\times
	(2p_1)(2p_2)\ \left\langle
	{}^i_rq_{12}(p_1)\, {}^i_rq^*_{12}(p_1)\,
	{}^j_sq_{34}(p_2)\, {}^j_sq^*_{34}(p_2)
	\right\rangle^{(\!2\!)}.
	\nonumber\\
\end{eqnarray}
When applying Wick's theorem to the four-point correlation function
$\langle q q^* q q^*\rangle^{(\!2\!)}$ there are two ways of pairing $q$ with $q^*$:
\begin{eqnarray}
	\left\langle
	{}^i_rq_{12}(p_1)\, {}^i_rq^*_{12}(p_1)\,
	{}^j_sq_{34}(p_2)\, {}^j_sq^*_{34}(p_2)
	\right\rangle^{(\!2\!)} &&
	\nonumber\\ && \hspace{-120pt} 
	= {}^i_rM^{-1}_{12,12}(p_1)\ {}^i_rM^{-1}_{34,34}(p_2)
	\nonumber\\ && \hspace{-100pt} 
	+ \left[M^{-1}_{12,34}(p_1)\right]^2\ \delta_{rs}\
	\delta_{ij}\ \delta_{p_1,p_2}.
\end{eqnarray}
Comparison with Eq.~(\ref{I}) shows that the first term on the right hand 
side corresponds to the disconnected part $(\langle I \rangle^{(2)})^2$.
The second term on the right hand side however yields a contribution
to the connected part $\langle I^2\rangle_c^{(\!2\!)}$.
As $M^{-1}_{12,34}(p)\sim\delta_{\alpha_1\alpha_3}\,\delta_{\alpha_2\alpha_4}$
it is again only the noninteracting part of $M^{-1}_{12,34}(p)$ that can 
give rise to a term proportional to $N_1 N_2$.

Again we can identify our result with a result obtained using standard 
perturbation theory.\cite{Cheung89}  Diffusons and cooperons can be shown
to give identical contributions.  It can be verified that we can write
\begin{equation}
	\langle I(\phi)^2 \rangle_c^{(\!2\!)} = \left.
	\frac{\partial^2\langle\Omega(\phi)\Omega(\phi')\rangle}
	{\partial\phi \partial\phi'}\right|_{\phi'=\phi},
\end{equation}
with
\begin{equation}
	\langle\Omega(\phi)\Omega(\phi')\rangle =
	\frac{4T}{\pi} \sum_p \sum_{\omega>0}
	\ln\left(\frac{1}{D p^2 + \omega}\right),
\end{equation}
where $\omega=2\pi Tn$, with integer $n\geq1$, 
and $p=(2\pi/L)$ $\times(n_p+(\phi+\phi')/\phi_0)$.
This is exactly the result obtained in Ref.~\onlinecite{Cheung89},
which at $T=0$ is given by
\begin{equation}
	\sqrt{\left\langle I(\phi)^2 \right\rangle_c^{(\!2\!)}} \simeq 
	I_D \left(\frac{2\sqrt{2}}{3\pi}\right)
	\sin\left[\frac{2\pi\phi}{\phi_0}\right].
\end{equation}
We stress again that this result is independent of the electron-electron
interaction.  Numerically, it falls short of explaining the experiment of
Ref.~\onlinecite{Chandrasekhar91} by one or two orders of magnitude.

\subsection{Persistent current distribution is Gaussian}
\label{sec:distribution}

In this section we will show that the higher moments of the current distribution, 
$\langle I^n\rangle^{(\!2\!)}$ with $n\geq3$, are exactly those of a Gaussian
distribution, provided that we use only the Gaussian part, $S^{(\!2\!)}[q]$,
of the action when evaluating them, i.e.\ we approximate $\langle\cdots\rangle$ 
by $\langle\cdots\rangle^{(2)}$.
In the remainder of this section we will suppress the Gaussian superscript
for notational convenience.

Let us start by observing that
\begin{eqnarray}
	e^{-S[q]} \frac{\partial^n}{\partial\phi_1\cdots\partial\phi_n}
	e^{S[q]} &=& 
	\nonumber\\ && \mbox{} \hspace{-100pt}
	\sum_{n_1,n_2\geq0} 
	\left(\frac{\partial S[q]}{\partial\phi_*}\right)^{n_1}
	\left(\frac{\partial^2 S[q]}
	{\partial\phi_*\partial\phi_*}\right)^{n_2}
	\delta_{n_1+2n_2,n}.
\end{eqnarray}
In this expression the fluxes $\phi_*$ denote a permutation of the
fluxes $\{\phi_i\}_{i=1}^n$.
The right hand side leads to correlation functions of the type
$\langle (qq^*)^{n_1} (qq^*)^{n_2}\rangle$,
which are factorized using Wick's theorem.

To facilitate the combinatorics of the problem it is helpful to introduce
a symbolic notation.
For the average current we have
\begin{equation}
	\langle I \rangle 
		= \langle \widehat{qq^*}\rangle
		= \langle \underline{\widehat{qq^*}}\rangle.
\end{equation}
The symbol $(\widehat{\cdots})$ corresponds to a differentiation with respect
to the flux, $\partial S/\partial\phi$, which produces a term $qq^*$; 
furthermore this symbol implies a term $(2p)(\partial p/\partial\phi)$
(see Eq.~(\ref{I})).
The line under $qq^*$ denotes a pairing of $q$ and $q^*$ that
produces a propagator $M^{-1}$.

For the second moment we have in symbolic notation
\begin{eqnarray}
	\langle I^2 \rangle 
	&=& \langle\widehat{qq^*}\widehat{qq^*\!}\rangle
	+ \langle\widehat{\widehat{qq^*\!}}\rangle
	\nonumber\\
	&=& \langle\underline{\widehat{qq^*\!}}\rangle
	\langle\underline{\widehat{qq^*\!}}\rangle
	+ \langle\underline{\widehat{qq^*\!}\widehat{qq^*\!}}\rangle
	+ \langle\underline{\widehat{\widehat{qq^*\!}}}\rangle.
\end{eqnarray}
On the last line of this equation, the first term corresponds to the disconnected
part of $\langle I^2\rangle$ which is just the square of the average current;
the remaining two terms constitute the connected part $\langle I^2\rangle_c$,
where the second term corresponds to
$(\partial S/\partial\phi_1)(\partial S/\partial\phi_2)$
and the third term corresponds to
$(\partial^2 S/\partial\phi_1\partial\phi_2)$.
Both contributions to $\langle I^2\rangle_c$ contain a factor
$(\partial p_1/\partial\phi_1)(\partial p_2/\partial\phi_2)$.
The first two terms arise from the two different ways in which $q$ can be paired
with $q^*$ in a product of the type $qq^*qq^*$, 
as discussed in Sec.~\ref{sec:Isq-Gaussian}.

To understand what happens for the moments $n\geq3$ it is important to 
recall that any wave number $p$ can depend on only two different fluxes
(see Sec.~\ref{sec:fluxes-wavenumbers}).
When two $q$'s are paired, their replica indices must be the same
since $M^{-1}_{\alpha_1\alpha_2\alpha_3\alpha_4}\sim%
\delta_{\alpha_1\alpha_2}\delta_{\alpha_3\alpha_4}$.
Consequently, when terms containing three or more fluxes are paired into a 
connected cluster, the corresponding expression will contain a term,
\begin{equation}
	\frac{\partial p}{\partial\phi_1} 
	\frac{\partial p}{\partial\phi_2} 
	\frac{\partial p}{\partial\phi_3} = 0,
\end{equation}
which vanishes by definition, since $p$ cannot depend on three fluxes.
In terms of the symbolic notation used here we have, e.g.,
\begin{equation}
	\langle\underline{\widehat{qq^*\!}\widehat{\widehat{qq^*\!}}}\rangle =
	\langle\underline{\widehat{qq^*\!}\widehat{qq^*\!}\widehat{qq^*\!}}\rangle
	= 0.
\end{equation}
It follows that when Wick's theorem is applied to a general term of the type
\begin{equation}
	\langle
	\widehat{qq^*} \cdots \widehat{qq^*}
	\
	\widehat{\widehat{qq^*\!}} \cdots \widehat{\widehat{qq^*\!}}
	\rangle,
\end{equation}
it factorizes into a product of connected terms that depend on either one or 
two fluxes.
There are three such terms, and all of them occur in the expressions
for $\langle I\rangle$ and $\langle I^2\rangle_c$.
Working out the combinatorics of the possible ways of pairing we obtain
\begin{eqnarray}\label{product}
	\langle I^n\rangle &=& 
	\sum_{{\scriptsize k,m\geq0}\atop{\scriptsize 2k+2m\leq n}}
	c^n_{km}
	\Big(\langle\underline{\widehat{qq^*\!}}\rangle\Big)^{n-2k-2m}
	\nonumber\\&&\mbox{}\times
	\Big(\langle\underline{\widehat{qq^*\!}
	\widehat{qq^*\!}}\rangle\Big)^k
	\Big(\langle\underline{\widehat{\widehat{qq^*\!}}}\rangle\Big)^m,
\end{eqnarray}
with
\begin{equation}\label{cnkl}
	c^n_{km} = \frac{n!}{k! \ m! \ (n-2k-2m)!} (\case{1}{2})^{k+m}.
\end{equation}
The characteristic function $\tilde p(z)$ is defined in terms of the persistent
current distribution function $p(I)$ as
\begin{equation}
	\tilde p(z) = \int_{-\infty}^\infty e^{zI} p(I)
	= \sum_{n=0}^\infty \frac{z^n}{n!} \langle I^n \rangle.
\end{equation}
Substituting Eqs.~(\ref{product}) and (\ref{cnkl}) for $\langle I^n\rangle$
into this expression we obtain
\begin{eqnarray}
	\tilde p(z) &=& \exp\Big[ 
	z \langle\underline{\widehat{qq^*\!}}\rangle
	+ \case{1}{2} z^2 \Big(
	  \langle\underline{\widehat{qq^*\!} \widehat{qq^*\!}}\rangle
	  + \langle\underline{\widehat{\widehat{qq^*\!}}}\rangle
	\Big)\Big] \nonumber\\
	&=& \exp\Big[ z \langle I\rangle + \case{1}{2} z^2 
	\langle I^2 \rangle_c \Big].
\end{eqnarray}
From this expression we read off that, as expected, the first two cumulants of 
the distribution are $\langle I\rangle$ and $\langle I^2\rangle_c$, respectively,
while all higher cumulants vanish.
This is the signature of the Gaussian distribution.

\section{Beyond Gaussian approximation}
\label{sec:beyond-Gaussian}

As we saw in the previous section, the only part of the action that depends
on the flux $\phi$ is the gradient term
\begin{equation}
	S_\nabla[Q] = \frac{-1}{2G} \int {\rm d}{\bf x}\
	{\rm tr}(\nabla Q({\bf x}))^2.
\end{equation}
In evaluating averages of the type $\langle qq^*\cdots qq^*\rangle$ 
we used the full Gaussian part $S^{(\!2\!)}[Q]$ of the action, 
i.e., all terms of order $O(q^2)$ including the interaction terms.
This is the reason why the Gaussian result for $\langle I\rangle$ can depend 
on the interaction coupling constant $K$, even if $S_\nabla[Q]$ does not.

In what follows we will take into account terms of higher order in the 
expansion in powers of $q$ of $S_\nabla[Q]$.
Averages $\langle qq^*\cdots qq^*\rangle$ however will still be evaluated
in Gaussian approximation with a weight factor $\exp(S^{(\!2\!)}[q])$, so that
Wick's theorem can be used to express the average in terms of products of
Gaussian propagators, as was done in the previous section in the case of
$\langle I^2\rangle$.
Since the disorder is relatively weak, it is expected that using the full
average $\langle\cdots\rangle$ instead of the Gaussian approximation 
$\langle\cdots\rangle^{(2)}$ would simply lead to small corrections to 
the terms computed here using the Gaussian average.

\subsection{Calculation of $\langle I\rangle^{(\!4\!)}$}

The first term in the expansion beyond Gaussian order in the expansion
of $S_\nabla[Q]$ in powers of $q$ is given by
\begin{eqnarray}
	S_\nabla^{(\!4\!)}[q] &=& \frac{-1}{2G(LL_\bot^2)^3}
	\sum_p \sum_{k_1,k_2} \sum_{1234} 
	\sum_{\{r_\alpha\}} \sum_{\{i_\alpha\}} p^2
	\\ \nonumber && \hspace{-20pt}\times\
	{}_{r_1}^{i_1}q_{12}(p+k_1)  \, {}_{r_2}^{i_2}q^*_{32}(k_1) \,
	{}_{r_3}^{i_3}q_{34}(-p+k_2) \, {}_{r_4}^{i_2}q^*_{14}(k_2)
	\nonumber\\ && \hspace{-20pt}\times\
	{\rm tr}(\epsilon_{r_1}\epsilon^\dagger_{r_2}
	         \epsilon_{r_3}\epsilon^\dagger_{r_4}) \,
	{\rm tr}(s_{i_1}s^\dagger_{i_2} s_{i_3}s^\dagger_{i_4}).
\end{eqnarray}
It should be noted that there is a very similar term in which the role of
$q$ and $q^\dagger$ is interchanged.  It can be verified that this
term yields an identical contribution, so that it is sufficient
to multiply the final result by a factor 2.
The quadruplets $(r_1r_2r_3r_4)$ for which
${\rm tr}(\epsilon_{r_1}\epsilon^\dagger_{r_2}\epsilon_{r_3}%
\epsilon^\dagger_{r_4})=1$ are
\begin{eqnarray*}
	& (0000) \quad (1111) \quad (2222) \quad (3333) & \\
	& (0220) \quad (1331) \quad (3113) \quad (2002) & \\
	& (0011) \quad (2233) \quad (3322) \quad (1100) & \\
	& (0231) \quad (1320) \quad (2013) \quad (3102) &
\end{eqnarray*}
while the trace vanishes for all other quadruplets.
Note that the allowed quadruplets all have the following property:
\begin{equation}
	\left[\begin{array}{c}-\\-\\+\\+\end{array}\right]_{r_1} 
	= \left[\begin{array}{c}-\\-\\+\\+\end{array}\right]_{r_4},
	\qquad
	\left[\begin{array}{c}-\\-\\+\\+\end{array}\right]_{r_2} 
	= \left[\begin{array}{c}-\\-\\+\\+\end{array}\right]_{r_3},
\end{equation}
and
\begin{equation}
	\left[\begin{array}{c}+\\-\\+\\-\end{array}\right]_{r_1} 
	= \left[\begin{array}{c}+\\-\\+\\-\end{array}\right]_{r_2},
	\qquad
	\left[\begin{array}{c}+\\-\\+\\-\end{array}\right]_{r_3} 
	= \left[\begin{array}{c}+\\-\\+\\-\end{array}\right]_{r_4}.
\end{equation}
The contribution of $S^{(\!4\!)}_\nabla[q]$ to the average current $\langle I\rangle$ 
reads
\begin{eqnarray}
	T \left\langle\frac{\partial S^{(\!4\!)}[q]}{\partial\phi}
	\right\rangle^{(\!2\!)} &=&
	T \sum_p \sum_{k_1,k_2} \sum_{1234} 
	\sum_{\{r_\alpha\}} \sum_{\{i_\alpha\}}
	(2p) \frac{\partial p}{\partial\phi}
	\nonumber\\ && \hspace{-80pt} \times \left\langle
	{}_{r_1}^{i_1}q_{12}(p+k_1)   {}_{r_2}^{i_2}q^*_{32}(k_1)
	{}_{r_3}^{i_3}q_{34}(-p+k_2) {}_{r_4}^{i_4}q^*_{14}(k_2)
	\right\rangle^{(\!2\!)},
	\nonumber\\
\end{eqnarray}
with the allowed values for $p$ given in Eq.~(\ref{p-phi}).
As in the preceding section, there are two ways of pairing $q$ with $q^*$.
Pairing $q_{12}(p+k_1)$ with $q^*_{32}(k_1)$ yields a Gaussian propagator
proportional to $\delta_{p,0}$, which together with the factor $2p$ leads to
a vanishing contribution.
The alternative pairing of $q_{12}(p+k_1)$ and $q^*_{32}(k_2)$ gives rise to 
the following expression:
\begin{eqnarray}
	\langle I \rangle^{(\!4\!)} &=&
	2T \sum_{k_1,k_2} \sum_{1234} 
	\sum_{r_1,r_2} \sum_{i_1,i_2}
	(k_2-k_1) \frac{\partial p}{\partial\phi}
	\nonumber\\ && \mbox{}\times
	{}_{r_1}^{i_1}M^{-1}_{1214}(k_2) {}_{r_2}^{i_2}M^{-1}_{3234}(k_1).
\end{eqnarray}
Note that $r_1=r_4$, $r_2=r_3$, $\alpha_2=\alpha_4$, and $n_2=n_4$.
The only contribution that survives in the replica limit is the one where 
all four replica labels are the same, which is only possible if for both 
propagators $M^{-1}$ we take into account the interacting part only,
as this part is completely diagonal in the replica labels.
We then have
\begin{eqnarray}\label{I4}
	\langle I \rangle^{(\!4\!)} &\simeq& \frac{T G^3 K^2_{\rm c}}{LL_\bot^2}
	\sum_{k_1,k_2}\
	\sum_{\omega_1,\omega_3>0}\ \sum_{\omega_2<0}\
	\sum_{r_1,r_2}
	(k_2-k_1) \frac{\partial p}{\partial\phi}
	\nonumber\\ && \mbox{}\times
	{\cal D}^2(k_2,\omega_1-\omega_2)
	{\cal D}^2(k_1,\omega_3-\omega_2).
\end{eqnarray}

To proceed it is necessary to analyze the dependence of $k_1$, $k_2$, and
$p$ on the flux $\phi$ and the indices $r_1$ and $r_2$.
In the Gaussian case we saw that only cooperons can contribute; here however
the situation is slightly more complicated.  We have
\begin{mathletters}
\begin{equation}
	p = \frac{2\pi}{L} 
	\left[n_p + \left(
	\left[\begin{array}{c}-\\-\\+\\+\end{array}\right]_{r_1} 
	- \left[\begin{array}{c}-\\-\\+\\+\end{array}\right]_{r_2} 
	\right) \frac{\phi}{\phi_0}\ \right].
\end{equation}
\begin{equation}
	k_1 = \frac{2\pi}{L} 
	\left(n_{k_1} + 
	\left[\begin{array}{c}0\\-\\+\\0\end{array}\right]_{r_2} 
	\frac{2\phi}{\phi_0}\ \right).
\end{equation}
\begin{equation}
	k_2 = \frac{2\pi}{L} 
	\left(n_{k_2} + 
	\left[\begin{array}{c}0\\-\\+\\0\end{array}\right]_{r_1} 
	\frac{2\phi}{\phi_0}\ \right).
\end{equation}
\end{mathletters}
From these relations it can be seen that the allowed 
quadruplets of the form $(r_1r_2r_2r_1)$ fall into three classes:
(i) $\partial p/\partial\phi=-2$ for $(r_1r_2)=(02)$ or (13);
(ii) $\partial p/\partial\phi=+2$ for $(r_1r_2)=(20)$ or (31);
(iii) $\partial p/\partial\phi=0$ for $r_1=r_2$.
It can be seen that the pairs (02), (20), (13), and (31) all give identical
contributions to $\langle I\rangle^{(\!4\!)}$.
For example, the contributions of (02) and (13) are equal since the expression
for $\langle I\rangle^{(\!4\!)}$ is invariant under the transformation:
$k_1\to -k_2$, $k_2\to -k_1$.

We are now ready to evaluate Eq.~(\ref{I4}).
Focusing on the case $(r_1r_2)=(02)$ and introducing a rescaled frequency 
$y=(\omega/D)(L/2\pi)^2$
we obtain
\begin{equation}
	\langle I(\phi)\rangle^{(\!4\!)} = 
	\frac{I_D(N_{\!F}\bar V)^2 L}{6\pi N_{\!F}DL_\bot^2} 
	F(\phi).
\end{equation}
We have written the prefactor of $F(\phi)$ in a way that allows for an easy 
comparison with the Gaussian order result, but it should be noted that this 
expression does not depend on the strength of the disorder, as can be seen 
by using $D=v_F\ell/3$.
The flux dependence is contained in the dimensionless expression,
\begin{eqnarray}\label{F-phi}
	F(\phi) &=& 
	\int_0^\infty dy
	\sum_{n_2=-\infty}^\infty
	\frac{n_2+2\phi/\phi_0}{(n_2+2\phi/\phi_0)^2+y}
	\sum_{n_1=1}^\infty
	\frac{2}{n_1^2 + y}
	\nonumber\\
	&=& 2\pi \int_0^\infty ds
	\frac{\sin 4\pi\phi}{\cosh 2s - \cos 4\pi\phi}
	\left(\coth s - \frac{1}{s}\right),
\end{eqnarray}
where we have used a change of variable to $s=\pi\sqrt{y}$.
The zero-momentum term $n_1=0$ has been excluded, the rationale for which
was given in Sec.~\ref{sec:zero-mode}.
The integral is convergent for all values of $\phi$;
this would not be the case if the zero-mode contribution had not been 
subtracted, since then there would be a small-frequency divergence.
We will now argue that the term $\langle I\rangle^{(4)}$ and all other 
higher-order terms, are small compared to the Gaussian contribution 
$\langle I\rangle^{(2)}$.

\subsection{Magnitude of general higher-order terms for $\langle I\rangle$}

In the preceding section we have performed the calculation of the first 
contribution to the average current beyond the Gaussian order.
We will now generalize this calculation to estimate the magnitude
of contributions of arbitrary order to $\langle I\rangle$ and $\langle I^2\rangle$.

Our aim is to write all expressions as a product of a dimensionful prefactor,
i.e., a combination of the parameters $D$, $L$, $L_\bot$, $N_{\!F}$, and $\bar V$,
and a dimensionless sum over rescaled wave numbers and frequencies.
Consider therefore the following contribution to the average current:
\begin{equation}
	\langle I\rangle^{(2m)} = -T \left\langle\frac{\partial S^{(\!2m\!)}[q]}
	{\partial\phi}\right\rangle.
\end{equation}
First, there is a factor $1/[G(L L_\bot^2)^{2m-1}]$, associated with
$S^{(\!2m\!)}[q]$.  Next, the terms $(2p)$ and $(\partial p/\partial\phi)$ each
yield a factor $(2\pi/L)$.
The differentiation of $S^{(\!2m\!)}[q]$ yields a $(2m)$-point correlation function
in terms of $q$;  applying Wick's theorem we obtain a product of $m$
Gaussian propagators $M^{-1}$.
Each propagator has a factor $GLL_\bot$ associated with it.

The propagators can either contribute their noninter\-acting 
or their interacting part; the latter is diagonal in the replica labels.
The number of interacting contributions is determined by the requirement 
that due to the replica limit there should only be a single independent
replica index.
This can only be achieved by choosing the interacting part of all propagators,
which is proportional to $[{\cal D}(k,\omega)]^2$.
There is a factor $(L/2\pi)^4$ coming from a rescaling of
$[{\cal D}(k,\omega)]^2$, and a factor $GK=N_{\!F}\bar V/D$ representing the
interaction.
For simplicity, we have neglected the logarithmic correction term that was
discussed in Sec.~\ref{sec:I-Gaussian}.
Collecting all factors we find that
\begin{equation}
	\langle I\rangle^{(2m)} \sim
	\frac{e(GLL_\bot^2)^m(T_0)^p}{G(LL_\bot^2)^{2m-1}}
	\left(\frac{2\pi}{L}\right)^2
	\left[\left(\frac{L}{2\pi}\right)^4(GK)\right]^m.
\end{equation}
The factor $(T_0)^p$, with $T_0=D/L^2$, comes from a rescaling of the frequency
variables.  Rather than by working out the structure of the frequency summations
in detail, we can determine $p$ by dimensional analysis.  
From the requirement that $\langle I\rangle$ be proportional to a frequency,
it follows that $p=m+1$, so that
\begin{equation}
	\langle I\rangle^{(2m)} \sim
	\frac{eD}{L^2} (N_{\!F}\bar V)^m
	\left[\left(\frac{L}{L_\bot}\right)^2\frac{1}{N_{\!F}DL} \right]^{m-1}
\end{equation}
Note that $m=1$ corresponds to the Gaussian result of Sec.~\ref{sec:I-Gaussian}.

To determine the value of the dimensionless combination $(L/L_\bot)^2/(N_{\!F}DL)$
we note that in three dimensions the density of states at the Fermi energy is given 
by $N_{\!F} = m k_{\!F}/\pi^2$, while $D=v_{\!F}\ell/3$,
so that $N_{\!F}DL = (k_{\!F}\ell)(k_{\!F}L)/3\pi^2\simeq 10^6$.
This is so large that in spite of the large factor 
$(L/L_\bot)^2 \simeq 10^4$ we have
\begin{equation}
	\left[(N_{\!F}V_{\rm eff})
	\left(\frac{L}{L_\bot}\right)^2\frac{1}{N_{\!F}DL}\right]
	\simeq 10^{-2}.
\end{equation}
Thus we conclude that the higher-order terms are small compared to the 
Gaussian result.

Although contributions thus seem to be smaller by a factor of about 
$10^{-2}$ for each higher order, a simultaneous rapid growth occurs
of the combinatorial factors arising, e.g., from the different ways of pairing
$q$'s when applying Wick's theorem.
The perturbation expansion in terms of the disorder therefore at best provides
us with an asymptotic series, a situation not unfamiliar in other contexts.

\subsection{Magnitude of general higher-order terms for $\langle I^2\rangle$}

The calculation of the magnitude of higher-order contributions to the second
moment $\langle I^2\rangle$ proceeds in a similar fashion.
As we have seen above in Sec.~\ref{sec:Isq-Gaussian} there are two types of
contributions.
First we have
\begin{equation}
	\langle I^2\rangle^{(2m)} = T^2
	\left\langle\frac{\partial^2 S^{(\!2m\!)}[q]}{\partial\phi\partial\phi'}
	\right\rangle.
\end{equation}
A difference with the calculation of $\langle I\rangle_{2m}$ is that the 
replica limit now requires the existence of two independent replica indices.
This can be achieved by choosing the noninteracting part of one of the
propagators, while using the interacting part of all $(m-1)$ others.
The noninteracting propagator contributes a factor $(L/2\pi)^2$ instead of
$(GK)(L/2\pi)^4$, so that
\begin{equation}
	\langle I^2\rangle^{(2m)} \sim
	\frac{e^2(GLL_\bot^2)^m(T_0)^p}{G(LL_\bot^2)^{2m-1}}
	\left[\left(\frac{L}{2\pi}\right)^4(GK)\right]^{m-1}.
\end{equation}
In this case the requirement that $\langle I^2\rangle$ be proportional to a 
frequency squared, again yields $p=m+1$, so that
\begin{equation}
	\langle I^2\rangle^{(2m)} \sim
	\frac{e^2D^2}{L^4} \left[(N_{\!F}\bar V)
	\left(\frac{L}{L_\bot}\right)^2\frac{1}{N_{\!F}DL} \right]^{m-1}.
\end{equation}
The second type of contribution to $\langle I^2\rangle$ is
\begin{eqnarray}
	\langle I^2\rangle^{(2m_1,2m_2)} &=&
	(eT)^2 \left\langle 
	\frac{\partial S^{(\!2m_1\!)}[q]}{\partial\phi}
	\frac{\partial S^{(\!2m_2\!)}[q]}{\partial\phi'}
	\right\rangle 
	\nonumber\\ && \hspace{-60pt}
	= \frac{e^2D^2}{L^4} (N_{\!F}\bar V)^{m_1+m_2-1}
	\left[\left(\frac{L}{L_\bot}\right)^2\frac{1}{N_{\!F}DL} 
	\right]^{m_1+m_2-2}.
	\nonumber\\
\end{eqnarray}
In this case the Gaussian result corresponds to $m=1$ and $m_1=m_2=1$, respectively.
Again we conclude that the corrections for $\langle I^2\rangle$ are small compared
to the Gaussian result.

\subsection{Convergence of the dimensionless sums}
\label{sec:sums}

In the preceding subsections we argued that all terms beyond Gaussian order
are small compared to the Gaussian result.
It is essential to verify that the corresponding dimensionless sums over 
wave numbers and frequencies are convergent.
Based on the structure of the calculation of $\langle I\rangle^{(4)}$ we
expect that any higher-order term $\langle I^n\rangle^{(2m)}$ will contain 
a dimensionless, flux-dependent part given by an expression of the type
\begin{equation}
	F_a^b(x)=\!\int_0^\infty\! dy
	\left( \sum_{n=-\infty}^\infty \frac{n+x}{(n+x)^2+y}\right)^a
	\left( \sum_{n=1}^\infty \frac{1}{n^2+y}\right)^b.
\end{equation}
Here $x$ denotes some linear combination of dimensionless fluxes $\phi_i/\phi_0$,
and the integers $a$ and $b$, satisfying $a+b=m$, are determined by the 
detailed structure of the allowed sets $\{r_i\}$ of particle-hole indices.
In the case of $\langle I\rangle^{(4)}$ we have $x=2\phi/\phi_0$ and $a=b=1$.
The integral over $y$ is always convergent for $y\to\infty$;
for generic values of $x$ there is also convergence for $y\to0$.
Only when $x$ is close to an integer a divergence may occur.
Let us restrict ourselves without loss of generality to values $|x|\ll 1$.
It is easy to show by a rescaling of $y$ that $F_a^b(x)\sim x^{2-a}$ as $x\to0$.
In other words, only when $a=1$ does the moment of the persistent current 
distribution under consideration go to zero in the limit of vanishing enclosed 
flux.
On the other hand, it is clear from symmetry considerations that $F_a^b(x)$ 
vanishes whenever $x$ is a multiple of $\case{1}{2}$.

In principle it is possible that the case $a\leq2$ never occurs, even at 
high orders in perturbation theory, but there is another reason why the apparent
discontinuity at $x=0$ should not be too alarming even if it would occur.
To see this, one should realize that the electrons are localized on a spatial
scale $\xi=\ell(k_FL_\bot)^2 \gg L$.
Values $|x|\ll 1$ correspond to length scales $L/x\gg\L$ that will be on the order
of $\xi$ if $x\sim(L/\ell)(k_FL_\bot)^{-2}$.
On this scale there no longer exists diffusive behavior.
In the context of the nonlinear $\sigma$ model diffusive behavior corresponds to
an ordered state breaking the symmetry of the model.
Goldstone modes and their associated diffusive propagators only occur in the
ordered state.
This means that for exactly those values of $x$ for which $F_a^b(x)$ diverges,
the nonlinear $\sigma$ model description of the quantum-mechanical problem breaks
down due to localization effects; the diffusive poles that are the source of
the divergence of $F_a^b(x)$ are replaced by correlation functions that are
exponentially damped in space, and the divergence disappears.
It should be noted that the above conceptual considerations are somewhat 
academic, in the sense that in the actual 
experiments\cite{Levy90,Chandrasekhar91,Mohanty96}
the temperature provides a natural cutoff for all possible divergences.

\section{Discussion}
\label{sec:discussion}

In this paper we have shown how a nonlinear $\sigma$ model approach to the
persistent current problem provides us with an efficient and unambiguous
method for calculating moments of the persistent current distribution for
a disordered ring.
We thus have developed a valuable new tool for studying the combined effect 
of quenched impurities and electron-electron interaction on the magnitude
of the persistent current at high orders in perturbation theory.
More traditional diagrammatic approaches so far have failed to provide a 
reliable method to analyze the consequences of high order interactions 
between slow modes.

Our main result is that higher-order terms order are small compared to
the lowest order Gaussian result (see Sec.~\ref{sec:beyond-Gaussian}).
Furthermore, we showed that the Gaussian approximation within the nonlinear
$\sigma$ model reproduces the diagrammatic results of 
Refs.~\onlinecite{Cheung89,Ambegaokar91,Eckern91}.
It follows that the experiments of Refs.~\onlinecite{Levy90,Chandrasekhar91} 
cannot be explained purely in terms of interacting Goldstone modes,
i.e., diffusons and cooperons, but that an additional theoretical 
ingredient is needed.

Using only the dominant Gaussian part of the action to calculate the 
higher moments of the persistent current distribution, we found
that this distribution is Gaussian (see Sec.~\ref{sec:distribution}).
This feature is not expected to survive in a more complete theory for
the tail of the distribution.

The nonlinear $\sigma$ model is essentially a small wave\-number and small 
frequency expansion.  However, it is known that ``nonhydrodynamic''
contributions may be important, an example being the density expansion of
transport coefficients for quantum systems.\cite{non-hydro}
It is therefore desirable to investigate the validity of the ``hydrodynamic'' 
approximation, by rederiving some of the known results without performing a 
small wave number expansion.
The relative importance of contributions coming from wave numbers in the
diffusive ($k\sim 1/L$) and ballistic ($k\sim 1/\ell$) range can then 
be determined.
An analysis along these lines will be presented in a separate publication.
\cite{B+K-future}

We believe that the method presented here can be used with some confidence, 
as in the simplest possible (Gaussian) approximation we exactly reproduce 
well-known and widely accepted results obtained using standard diagrammatic 
perturbation theory; this was shown in Sec.~\ref{sec:Gaussian}.
To illustrate the efficiency of our approach we note that the derivation of
the equivalent of our Gaussian results using diagrammatic perturbation theory 
requires three separate infinite resummations:
(i) in standard Born approximation the dressed Green's function 
contains an infinite number of impurity lines;
(ii) the diffusons and cooperons used as building blocks in the 
diagrammatic approach involve infinite ladder summations;
(ii) the logarithmic suppression of the electron-electron interaction in
standard perturbation theory is obtained by means of an infinite ladder 
summation of cooperons alternated with e-e interaction vertices.\cite{Eckern91}

An interesting observation is that, as opposed to the higher-order 
contributions, the Gaussian results are independent of the width $L_\bot$
of the ring, or alternatively the number of channels $(k_{\!F}L_\bot)^2$.
Although detailed calculations rapidly become more complicated beyond the
Gaussian approximation, it is relatively easy to give a naive estimate of 
the magnitude of higher-order terms (see Sec.~\ref{sec:beyond-Gaussian}).
It was tentatively concluded in Sec.~\ref{sec:beyond-Gaussian} that the
higher-order terms are small compared to the Gaussian results.

All calculations in this paper were performed using the grand canonical 
ensemble, in which the particle number fluctuates around an average value 
determined by the chemical potential.
Using the method of Altshuler {\em et al.},\cite{Altshuler91a} it would be in 
principle be possible to calculate corrections to the grand canonical results 
due to a fixed number of electrons in the ring.
In particular at higher order in perturbation theory, it would be interesting to
assess whether these corrections are negligible or not.

Our results indicate that although the Gaussian result is independent of
the thickness $L_\bot$ of the ring, the higher-order terms become more 
important as $L_\bot$ is decreased, if all other parameters are kept fixed.
At the same time however, decreasing $L_\bot$ brings the system closer to a 
one-dimensional system, in which electrons are localized and the description 
in terms of diffusive modes breaks down.
Thus it is hard to estimate the dependence of the persistent current on 
$L_\bot$, and it would be desirable to experimentally investigate it.

\acknowledgements

We would like to thank 
Dietrich Belitz, Bob Dorfman, Manher Jariwala,
Arnulf Latz, Raj Mohanty, and Masao Yoshimura
for stimulating discussions and comments.
This research was performed under NSF Grant No. DMR-96-32978.


\begin{references}
\bibitem{Hund}		F. Hund, Ann.\ Phys.\ (Leipzig) {\bf 32}, 102 (1938);
			{\bf 5}, 1 (1996).
\bibitem{Mailly95}	D. Mailly, C. Chapelier, and A. Benoit,
			Phys.\ Rev.\ Lett.\ {\bf 70}, 2020 (1993).
\bibitem{Levy90}	L. P. L\'evy, G. Dolan, J. Dunsmuir, and H. Bouchiat,
			Phys.\ Rev.\ Lett.\ {\bf 64}, 2074 (1990).
\bibitem{Chandrasekhar91}	
			V. Chandrasekhar, R. A. Webb, M. J. Brady, M. B. Ketchen,
			W. J. Gallagher, and A. Kleinasser,
			Phys.\ Rev.\ Lett.\ {\bf 67}, 2578 (1991).
\bibitem{Mohanty96}	P. Mohanty, E. M. Q. Jariwala, M. B. Ketchen, and R. A. Webb,
			in {\em Quantum Coherence and Decoherence},
			edited by K. Fujikawa and Y. A. Ono
			(Elsevier, New York, 1996).
\bibitem{clean-theory}	N. Byers and C. N. Yang, 
			Phys.\ Rev.\ Lett.\ {\bf 7}, 46 (1961);
			M. B\"uttiker, Y. Imry, and R. Landauer,
			Phys.\ Lett.\ {\bf 96A}, 365 (1983);
			R. Landauer and M. B\"uttiker,
			Phys.\ Rev.\ Lett.\ {\bf 54}, 2049 (1985).
\bibitem{exact-diag}	R. Berkovits and Y. Avishai, 
			Europhys.\ Lett.\ {\bf 29}, 475 (1995);
			G. Bouzerar and D. Poilblanc,
			Phys.\ Rev.\ B {\bf 52}, 10772 (1995);
			G. Chiappe, J. A. Verg\'es, and E. Louis,
			Solid State Commun.\ {\bf 99}, 717 (1996). 
\bibitem{hartree-fock}	H. Kato and Y. Yoshioka,
			Physica B {\bf 212}, 251 (1995).
\bibitem{Bouchiat89}	H. Bouchiat and G. Montambaux,
			J. Phys. (France) {\bf 50}, 2695 (1989).
\bibitem{Cheung89}	H. F. Cheung, E. K. Riedel, and Y. Gefen,
			Phys.\ Rev.\ Lett.\ {\bf 62}, 587 (1989).
\bibitem{Ambegaokar91}	V. Ambegaokar and U. Eckern, 
			Phys.\ Rev.\ Lett.\ {\bf 65}, 381 (1991).
\bibitem{Eckern91}	U. Eckern, Z. Phys.\ B {\bf 82}, 393 (1991).
\bibitem{Eckern92}	U. Eckern and A. Schmid,
			Europhys.\ Lett.\ {\bf 18}, 457 (1992).
\bibitem{Smith92}	R. A. Smith and V. Ambegaokar,
			Europhys.\ Lett.\ {\bf 20}, 161 (1992).
\bibitem{Beal-Monod92}	M. T. B\'eal-Monod and G. Montambaux,
			Phys.\ Rev.\ B {\bf 46}, 7182 (1992).
\bibitem{Kopietz93}	P. Kopietz, Phys.\ Rev.\ Lett.\ {\bf 70}, 3123 (1993).
\bibitem{Vignale94}	G. Vignale, Phys.\ Rev.\ B {\bf 50}, 7668 (1994).
\bibitem{Mueller94}	A. M\"uller-Groeling and H. Weidenm\"uller
			Phys.\ Rev.\ B {\bf 49}, 4752 (1994).
\bibitem{Schmid91}	A. Schmid, Phys.\ Rev.\ Lett.\ {\bf 66}, 80 (1991).
\bibitem{VonOppen91}	F. von Oppen and E. Riedel,
			Phys.\ Rev.\ Lett.\ {\bf 66}, 84 (1991).
\bibitem{Altshuler91a}	B. L. Altshuler, Y. Gefen, and Y. Imry,
			Phys.\ Rev.\ Lett.\ {\bf 66}, 88 (1991).
\bibitem{Abrikosov}	A. A. Abrikosov, L. P. Gorkov, and I. E. Dzyaloshinsky,
			{\em Methods of Quantum Field Theory in Statistical Physics}
			(Dover, New York, 1975).
\bibitem{Smith-thesis}	R. A. Smith, Ph.\ D. thesis, Cornell University, 1993.
\bibitem{Wegner}	F. Wegner, Z. Phys. {\bf 35}, 207 (1979).
\bibitem{Finkelstein}	A. M. Finkel'stein, Zh.\ Eksp.\ Teor.\ Fiz.\ {\bf 84},
			168 (1983) [Sov.\ Phys.\ JETP {\bf 57}, 97 (1983)].
\bibitem{revmodphys}	D. Belitz and T. R. Kirkpatrick, 
			Rev.\ Mod.\ Phys.\ {\bf 66}, 261 (1994).
\bibitem{Thouless}	D. J. Thouless, Phys.\ Rev.\ Lett.\ {\bf 39}, 1167 (1977).
\bibitem{Altshuler91b}	B. L. Altshuler, V. E. Kravtsov, and I. V. Lerner, 
			in {\em Mesocopic Phenomena in Solids}, edited by
			B. L. Altshuler, P. A. Lee, and R. A. Webb
			(Elsevier, Amsterdam, 1991).
\bibitem{Zinn-Justin}	J. Zinn-Justin, {\em Quantum Field Theory and Critical
			Phenomena} (Clarendon, Oxford, 1989).
\bibitem{non-hydro}	T. R. Kirkpatrick and J. R. Dorfman,
			Phys.\ Rev.\ A {\bf 22}, 1022 (1983);
			T. R. Kirkpatrick and D. Belitz,
			Phys.\ Rev.\ B {\bf 34}, 2168 (1986).
\bibitem{B+K-future}	H. J. Bussemaker and T. R. Kirkpatrick (unpublished).
\end{references}
\end{document}